\documentclass[aps, prd, nofootinbib, amsmath, twocolumn, superscriptaddress, preprintnumbers]{revtex4-1}
\usepackage{amssymb,esvect,amsmath,graphicx,latexsym,amsthm,slashed,eso-pic,hyperref}
\DeclareMathOperator{\ev}{eV} \DeclareMathOperator{\kev}{keV}  \DeclareMathOperator{\gev}{GeV}  \DeclareMathOperator{\cm}{cm}   \DeclareMathOperator{\km}{km}  \DeclareMathOperator{\s}{s}     
        \newcommand{\cO}{{\cal O}}

\newcommand{\pL}{\left(} \newcommand{\pR}{\right)} \newcommand{\bL}{\left[} \newcommand{\bR}{\right]} \newcommand{\cbL}{\left\{} \newcommand{\cbR}{\right\}}  
\newcommand{\beq}{\begin{equation}} \newcommand{\eeq}{\end{equation}}
\newcommand{\bea}{\begin{eqnarray}} \newcommand{\eea}{\end{eqnarray}}
\newcommand{\alg}[1]{\begin{align} \begin{split} #1 \end{split}  \end{align}}
\newcommand{\vev}[1]{\langle {#1} \rangle}
\newcommand{\tenx}[1]{\times 10^{#1}}
\newcommand{\Eq}[1]{Eq.~(\ref{#1})}  
  
\newcommand{\Fig}[1]{Fig.~\ref{#1}}

\usepackage{ulem}

\newcommand{\Neff}{N_{\rm eff}}
\newcommand{\DNeff}{\Delta \Neff}
\newcommand{\DNeffmax}{\DNeff^{\rm max}}
\newcommand{\Nnu}{N_\nu}

\newcommand{\zmr}{z_{\rm mr}}
\newcommand{\gsim}{\lower.7ex\hbox{$\;\stackrel{\textstyle>}{\sim}\;$}}
\newcommand{\lsim}{\lower.7ex\hbox{$\;\stackrel{\textstyle<}{\sim}\;$}}

\begin{document}
\title{Cosmologically Degenerate Fermions}
\author{Marcela Carena}
\email{carena@fnal.gov}
\affiliation{Fermi National Accelerator Laboratory, Batavia,  Illinois, 60510, USA}
\affiliation{Enrico Fermi Institute, University of Chicago, Chicago, Illinois, 60637, USA}
\affiliation{Kavli Institute for Cosmological Physics, University of Chicago, Chicago, Illinois, 60637, USA}
\affiliation{Department of Physics, University of Chicago, Chicago, Illinois, 60637, USA}
\author{Nina M. Coyle}
\email{ninac@uchicago.edu}
\affiliation{Enrico Fermi Institute, University of Chicago, Chicago, Illinois, 60637, USA}
\affiliation{Department of Physics, University of Chicago, Chicago, Illinois, 60637, USA}
\author{Ying-Ying Li}
\email{yingying@fnal.gov}
\affiliation{Fermi National Accelerator Laboratory, Batavia,  Illinois, 60510, USA}
\author{Samuel D.~McDermott}
\email{samueldmcdermott@gmail.com}
\affiliation{Fermi National Accelerator Laboratory, Batavia,  Illinois, 60510, USA}
\author{Yuhsin Tsai}
\email{ytsai3@nd.edu}
\affiliation{Department of Physics, University of Notre Dame, South Bend, IN 46556, USA}
\date{\today}

\begin{abstract}
Dark matter (DM) with a mass below a few keV must have a phase space distribution that differs substantially from the Standard Model particle thermal phase space: otherwise, it will free stream out of cosmic structures as they form.
We observe that {\it fermionic} DM $\psi$ in this mass range will have a non-negligible momentum in the early Universe, even in the total absence of thermal kinetic energy. This is because the fermions were inevitably more dense at higher redshifts, and thus experienced Pauli degeneracy pressure. They fill up the lowest-momentum states, such that a typical fermion gains a momentum $\sim \cO(p_F)$ that can exceed its mass $m_\psi$. We find a simple relation between $m_\psi$, the current fraction $f_\psi$ of the cold DM energy density in light fermions, and the redshift at which they were relativistic. Considering the impacts of the transition between nonrelativistic and relativistic behavior as revealed by measurements of $\DNeff$ and the matter power spectrum, we derive qualitatively new bounds in the $f_\psi-m_\psi$ plane.
We also improve existing bounds for $f_\psi = 1$ by an order of magnitude to $m_\psi=2$ keV. We remark on implications for direct detection and suggest models of dark sectors that may give rise to cosmologically degenerate fermions.
\end{abstract}
\preprint{FERMILAB-PUB-21-325-T}
\maketitle%

\noindent
{\bf Introduction.} As searches for canonical weak-scale dark matter (DM) candidates return null results, novel theoretical possibilities for the identity of the DM are gaining unprecendented traction \cite{Schumann:2019eaa}. 
Driven by these pressures from experiments, searches for ``light'' DM particles are entering a new and productive phase \cite{Lin:2019uvt}. 

Conventional wisdom provides several definitions for the dividing mass below which DM particles are ``light''. First is the operational definition driven by the fact that conventional direct detection searches with cryogenic materials suffer from poor kinematics for DM masses below a few GeV.  Next is the definition from considering the growth of structure that can be divined from studying the Cosmic Microwave Background (CMB) or Universe's structures larger than $\mathcal{O}(10)$~kpc:
if thermally produced DM is less massive than a few keV, the observed CMB and matter power spectrum will be modified. Finally, one can define light DM particles as those which are necessarily in a high occupation mode in the Milky Way (MW) today: DM is considered light if its de Broglie wavelength satisfies $m_{\rm DM} v_{\rm DM,\, MW} \lesssim (\rho_{\rm DM,\, MW}/m_{\rm DM})^{1/3}$,
which is true for $m_{\rm DM} \lesssim \cO(10)\ev$.
An interesting corollary to this final statement holds for fermions, which obey the Pauli exclusion principle \cite{1925ZPhy...31..765P} 
and thus are endowed with a ``Fermi momentum'', $p_F = (6\pi^2 n_\psi/g_\psi)^{1/3}$ for a fermion $\psi$ with $g_\psi$ internal degrees of freedom: if the DM is fermionic and lighter than $\sim \cO(10)$ eV, the Pauli degeneracy pressure could ``crowd it out'' of the Milky Way.

Extending these final considerations to cosmological scales will be the focus of this work. For clarity, we will explicitly enumerate some of the assumptions that will allow us to extrapolate the physics of degenerate fermion systems to different temperature and density scales than have been considered before. We assume that: 
\begin{itemize}
\item {\it the DM is thermally cold}: the $\psi$ particles have negligible random thermal motion, so that their kinetic energy comes solely from their degeneracy
\item {\it the DM has a fixed comoving number since the beginning of the Big-Bang Nucleosynthesis (BBN) until today}: the DM does not annihilate away. The DM can either be asymmetric or its annihilation is kinematically forbidden
\item {\it the DM does not form bound states in high-density environments}: such as confinement via nonabelian symmetry since the beginning of the BBN
\item {\it there is a single ``flavor'' of $\psi$}: we will assume $g_\psi=2$ for most of the discussion, appropriate for a single spin-1/2 degree of freedom. If we allow $N_f$ flavors of $\psi$ particles, then our constraints on $m_\psi$ weaken such that $m_\psi N_f^{1/4}$ remains constant.
\end{itemize}
Furthermore, we will consider the possibility that only a subdominant component of the present-day ($z=0$) dark matter density is in the form of $\psi$ particles. We will parameterize this fraction by the constant $f_\psi \equiv \rho_\psi/(\Omega_{\rm DM} \rho_c)$, where $\rho_\psi$ is the present-day energy density of $\psi$ particle, $\Omega_{\rm DM} \simeq 0.25$ is the fraction of the present-day energy density of the Universe in dark matter, and $\rho_c = 3H_0^2/8\pi G$ is the present-day critical density. When we consider $f_\psi<1$, we imagine that the remaining $1-f_\psi$ of today's DM behaves like a conventional cold DM (CDM) at all relevant times.

In this work, we provide constraints on light fermionic DM (or a subcomponent thereof) by considering the fact that the DM number density was higher at larger redshift, according to $n_\psi \propto (1+z)^3$. The Fermi momentum correspondingly scales like $p_F \propto (1+z)$. Thus, at some redshift, $z_t$, the Fermi momentum will satisfy $p_F(z_t)=m_\psi$. At $z_t$ and higher redshifts, we have $p_F \geq m_\psi$, in which case the typical $\psi$ particle is relativistic, since the average momentum $\vev{p} = 3p_F/4$ for a degenerate fermion gas. The fermionic DM becomes relativistic dark radiation at $z\geq z_t$, even in the absence of a dark-sector temperature. The fermionic dark ``matter" therefore suffers two types of constraints: cosmological constraints on the presence of extra radiation, and a Tremaine-Gunn-like bound~\cite{PhysRevLett.42.407} for non-thermalized neutral fermions that can modify the  structure formation process at $z> z_t$.
\\
\\
{\bf Cosmic Degeneracy Pressure and $\Delta N_{\rm eff}$ constraints.}
As discussed above, for a given present-day energy density $\rho_\psi$ dominated by the non-relativistic particle's rest energy, there will inevitably be a redshift $z_t$ where the Fermi momentum becomes comparable to the particle mass: $p_F(z_t) \equiv m_\psi$. Given the $\psi$ particle number density $n_\psi(z) =n_\psi(z=0) (1+z)^3$ with $n_\psi(z=0) = \frac{\rho_\psi}{N_f m_\psi}$, we obtain
\alg{ \label{eq:zt}
1+z_t =  \left[\frac{g_\psi N_f m_\psi^{4}}{6 \pi^2 f_\psi \Omega_{\rm DM} \rho_c}\right]^{1/3} \simeq \frac{1500}{ f_\psi^{1/3}} \pL\frac{ m_\psi}\ev \pR^{4/3}\,.
}
The second expression comes from assuming $g_{\psi}=2$ and $N_f=1$; these will be our default values throughout. 
At $z>z_t$, the DM particle redshifts like radiation as long as it satisfies the assumptions laid out in the introduction. We discuss the non-relativistic to relativistic transition in more detail in Appendix \ref{ap:thermo}, where we calculate the equation of state $w$ and show that $p_F \sim m_\psi$ is a reasonable estimate of this transition. 

\begin{figure}[t]
\begin{center}
\includegraphics[width=0.486\textwidth]{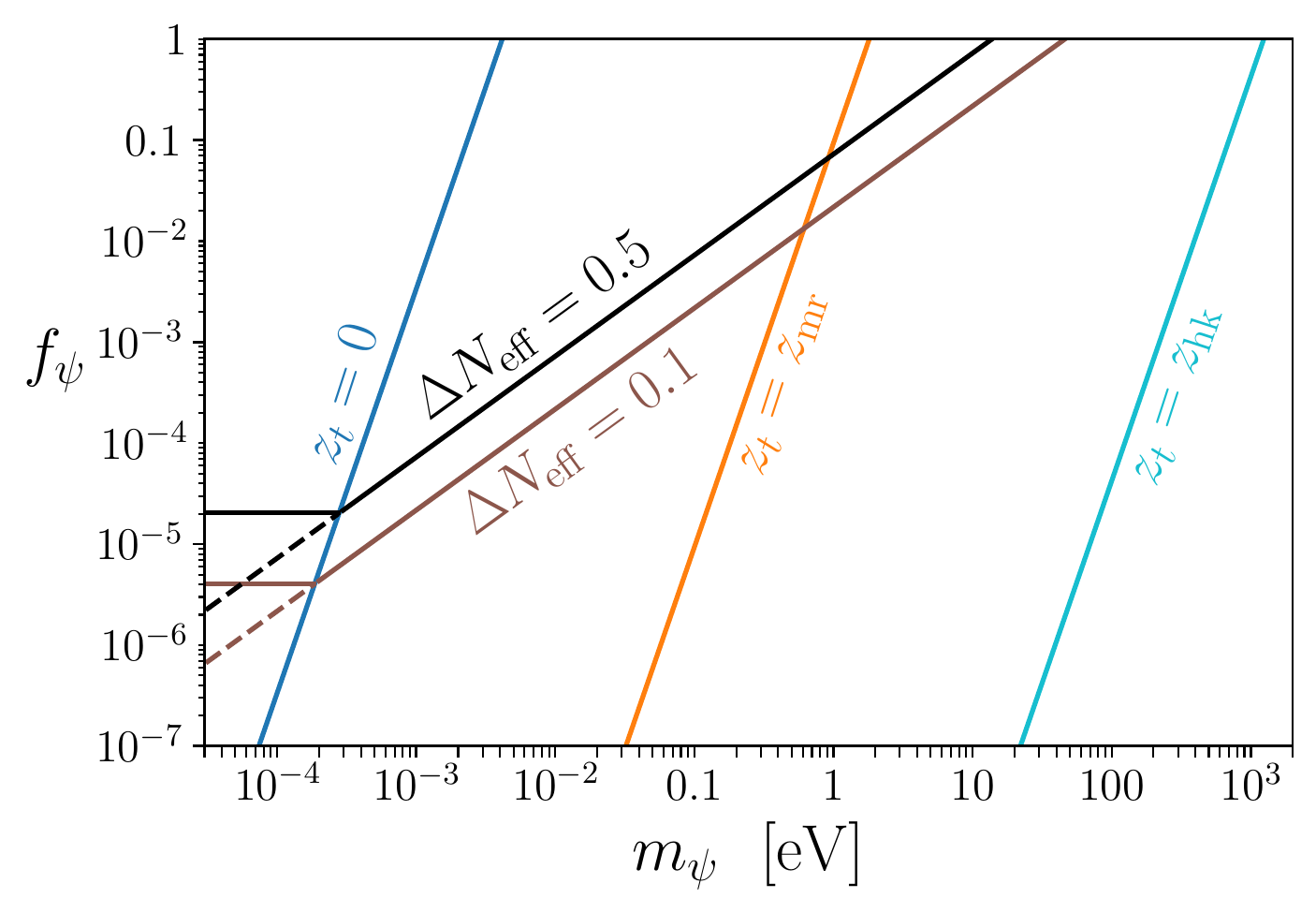}
\caption{Contours of $z_t$ as in Eq.~\ref{eq:zt} and of $\DNeff$ as in \Eq{eq:dneff}. We define $z_{\rm mr}=3300$ and $z_{\rm hk}=2\times10^7$.}
\label{fig:contours}
\end{center}
\end{figure}

Because the energy density in matter redshifts differently than the energy density in radiation, the relative value of the energy density of $\psi$ compared to the conventional radiation energy density will change as a function of $z$, up to the redshift $z_t$. We will characterize the energy density of $\psi$ over the redshift range for which it is relativistic 
by its equivalent number of effective neutrino degrees of freedom, $\DNeff$. 
We find that at $z>z_t$, 
\beq \label{eq:dneff}
\DNeff(m_\psi,f_\psi) = \frac{f_\psi(1+ \kappa \Nnu)}\kappa \frac{1+ \zmr}{1+ z_t(m_\psi,f_\psi)},
\eeq
where we have defined a constant  $\kappa=\frac78(\frac4{11})^{4/3}=0.22$, and the Standard Model number of neutrinos $\Nnu = 3.045$.
We plot contours of fixed values of $z_t$ and of $\DNeff(m_\psi, f_\psi)$ in the $f_\psi-m_\psi$ plane in Fig.~\ref{fig:contours}. Along the dashed portions of the lines, $z_t$ is negative, which leads to unphysical values of $\DNeff$.

\begin{figure}[t]
\begin{center}
\includegraphics[width=0.486\textwidth]{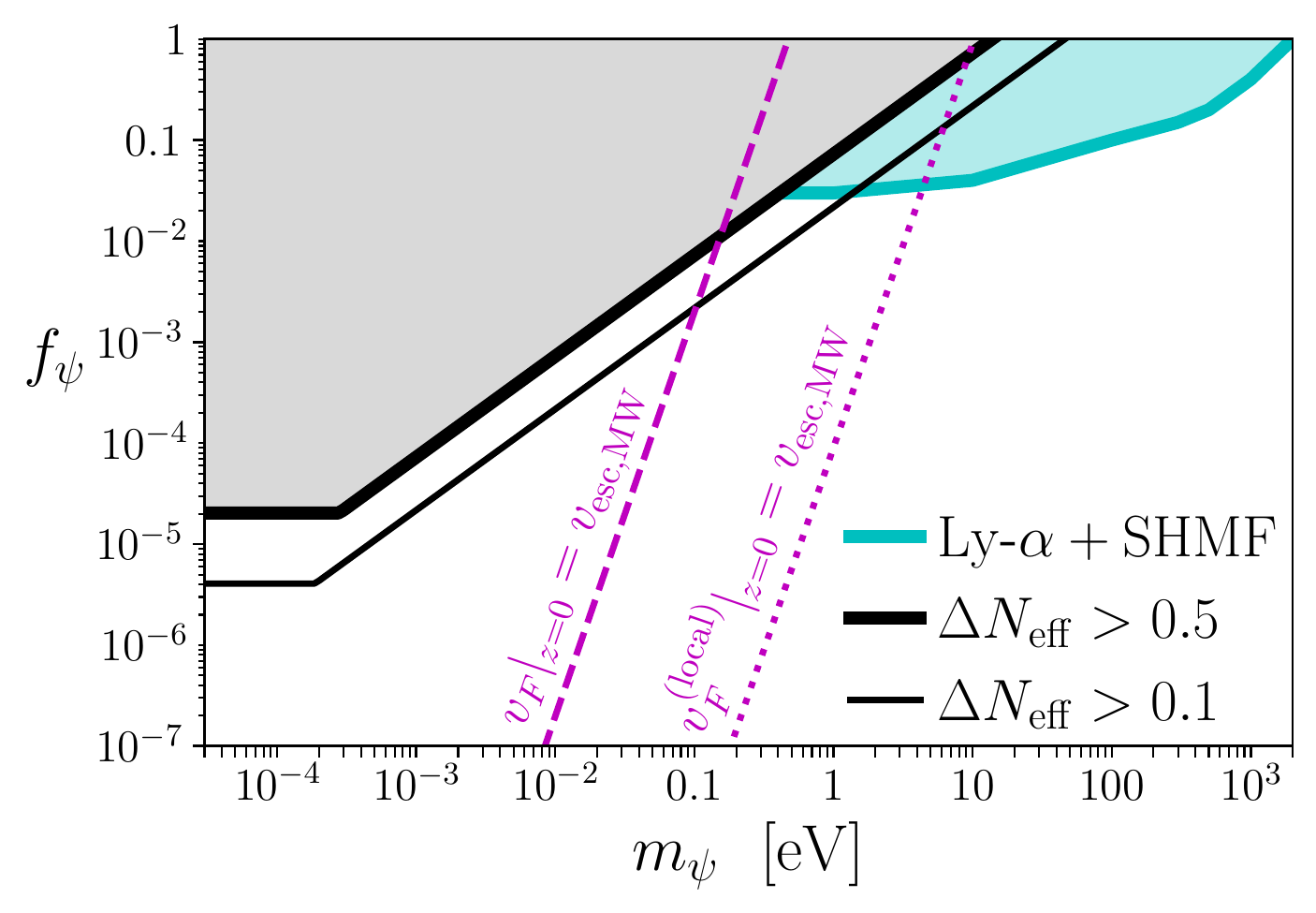}
\caption{Shaded regions are excluded due to fermion degeneracy. The energy density in the $\psi$ fluid in the early Universe exceeds the bound from measurements of $\DNeff$ \Eq{Neff-bound} (black). The novel redshift dependence of the $\psi$ fluid diminishes the matter power spectrum as measured by the Lyman-$\alpha$ forest and the subhalo mass function of the MW (cyan). The velocity due to degeneracy pressure alters the phase-space distribution of $\psi$ particles in the MW (magenta), due to the smooth background density (dashed) or due to the local Milky-Way based overdensity (dotted).}
\label{exclusion}
\end{center}
\end{figure}

One observable way in which $\rho_\psi$ can pose a problem is if the amount of extra radiation energy that $\psi$ carries at $z\geq z_t$ exceeds the allowed energy density from cosmological constraints; we will label this maximum allowed energy density in terms of the corresponding $\DNeffmax(z)$. Requiring $\DNeff$ in \Eq{eq:dneff} to be smaller than $\DNeffmax(z)$ and using \Eq{eq:zt} to replace $z_t$, we have the bound  
\beq
f_\psi <  \bL \frac{\DNeffmax(z)}{0.1} \bR^{3/4} \frac{m_\psi}{46 \ev}.
\eeq
Note that this bound only applies if $\rho_\psi\geq\kappa \DNeffmax(z) \Omega_\gamma \rho_c$ ($f_\psi \geq f_{\rm min} \equiv 4\tenx{-6} \DNeffmax(z)/0.1$). (We should also check that $z_t$ does not exceed the value of $z$ from which we extract the bound on $\DNeff$, which we discuss presently.) Summarizing, we have the bound
\beq \label{Neff-bound}
f_\psi < {\rm Max}\cbL f_{\rm min}, \, \bL \frac{\DNeffmax(z)}{0.1} \bR^{3/4} \frac{m_\psi}{46 \ev} \cbR.
\eeq
The values of $f_{\rm min}$ are plotted as the horizontal lines in Fig.~\ref{fig:contours}.
In a single-parameter extension of the baseline $\Lambda$CDM model, the constraint is $\DNeffmax (z \simeq z_{\rm CMB}) \simeq 0.28$ at 95\% CL due to CMB temperature and polarization measurements plus a prior on the baryon density from BAO, where $z_{\rm CMB}=10^3$ \cite{Aghanim:2018eyx}. 
A BBN-only calculation using the latest value of the D($p,\gamma)^3{\rm He}$ rate \cite{Mossa:2020gjc} provides the independent constraint $\DNeffmax (z\simeq z_{\rm BBN}) \simeq 0.12$~\cite{Yeh:2020mgl} from fitting the free parameter $\DNeff$ to the observed abundances of baryons, deuterium, and helium, where $z_{\rm BBN}\simeq 3\times10^9$ is the temperature of $n-p$ freezeout \cite{Berlin:2019pbq, Grohs:2016vef}. This is relaxed to $\DNeffmax (z \simeq z_{\rm BBN}) \simeq 0.37$ from a joint fit of the deuterium and helium abundances only \cite{Mossa:2020gjc}. Multiple-parameter extensions of $\Lambda$CDM, particularly those that reduce the Hubble tension, broaden the posteriors on all parameters, and the constraint is relaxed to $\DNeffmax \simeq 0.5$ \cite{Aghanim:2018eyx}.
We plot \Eq{Neff-bound} with $\DNeffmax=0.5$ or 0.1 for all $z$ in black in \Fig{exclusion}. The fact that we use a bound from BBN justifies the assumption that $z_t$ is smaller than the value of $z$ from which we extract the bound on $\DNeff$.
\\
\\
{\bf Structure formation constraints.} 
To this point, we have considered the impact of changing the {\it radiation} energy density of the Universe, but we may also consider the impacts of changing the CDM density. If an order one fraction of today's $\rho_{\rm CDM}$ remains relativistic below the redshift $z_{\rm hk} \simeq 2\tenx7$ when the high-$k$ modes in the Large Scale Structure, high-$\ell$ modes in the CMB, or galaxies with sizes $k^{-1}\gtrsim\mathcal{O}(10)$~kpc begin to form, there can be observable consequences. As discussed below, these consequences are revealed at low redshift by measurements of the matter power spectrum at different characteristic wavenumbers.

If $z_t \gtrsim z_{\rm hk}$, $\psi$ is cold for purposes of structure formation, and will be an indistinguishable part of the general CDM density. Thus, the cyan line in Fig.~\ref{fig:contours} suggests that the matter power spectrum is the same as $\Lambda$CDM for any value of $f_{\psi}$ if $m_\psi \gtrsim \cO(1)$ keV. If on the other hand $z_t \lesssim z_{\rm hk}$, $\psi$ is warm due to the degeneracy pressure and does not clump to form structures sufficiently early. This slows down the overall growth of matter density perturbations. The effect can be constrained, as we discuss presently, by the Lyman-$\alpha$ forest data or by counting MW satellite galaxies to determine the sub-halo mass function (SHMF).

A detailed simulation of the non-linear physics involved in the formation of the Lyman-$\alpha$ forest or the collapse of small-scale halos is beyond the scope of this work. Instead, we compare the linear power spectrum $P_{\psi}(k)$ for degenerate fermions to the results from the warm DM (WDM) scenarios that saturate the bound obtained in~\cite{Murgia:2017lwo}. We use the momentum distribution with large chemical potential in Appendix~\ref{ap:distributions} to mimic the momentum distribution of degenerate $\psi$ particles. Using the non-cold DM module of \texttt{CLASS}~\cite{Blas_2011} and the default $\Lambda$CDM parameters based on~\cite{Aghanim:2018eyx}, we calculate the linear matter power spectrum $P_{\psi}(k)$ for a given $f_{\psi}$ and $m_\psi$. Normalizing $P_{\psi}(k)$ to that of a $\Lambda$CDM Universe augmented by the presence of the same $\DNeff$ as obtained from \Eq{eq:dneff} gives the transfer function $T^2(k)\equiv P_{\psi}(k)/P_{\Lambda{\rm CDM}+\DNeff}(k)$.
The calculation is done for $z=4.2$, which is close to the redshift of Lyman-$\alpha$ data from the MIKE/HIRES+XQ-100 combined dataset used in~\cite{Irsic:2017ixq,Murgia:2017lwo}. The result only changes mildly from $z=0$.
The $T^2(k)$ spectrum varies between models with different DM masses, density fractions, and momentum distributions.

\begin{figure}[t]
\begin{center}
\includegraphics[width=0.486\textwidth]{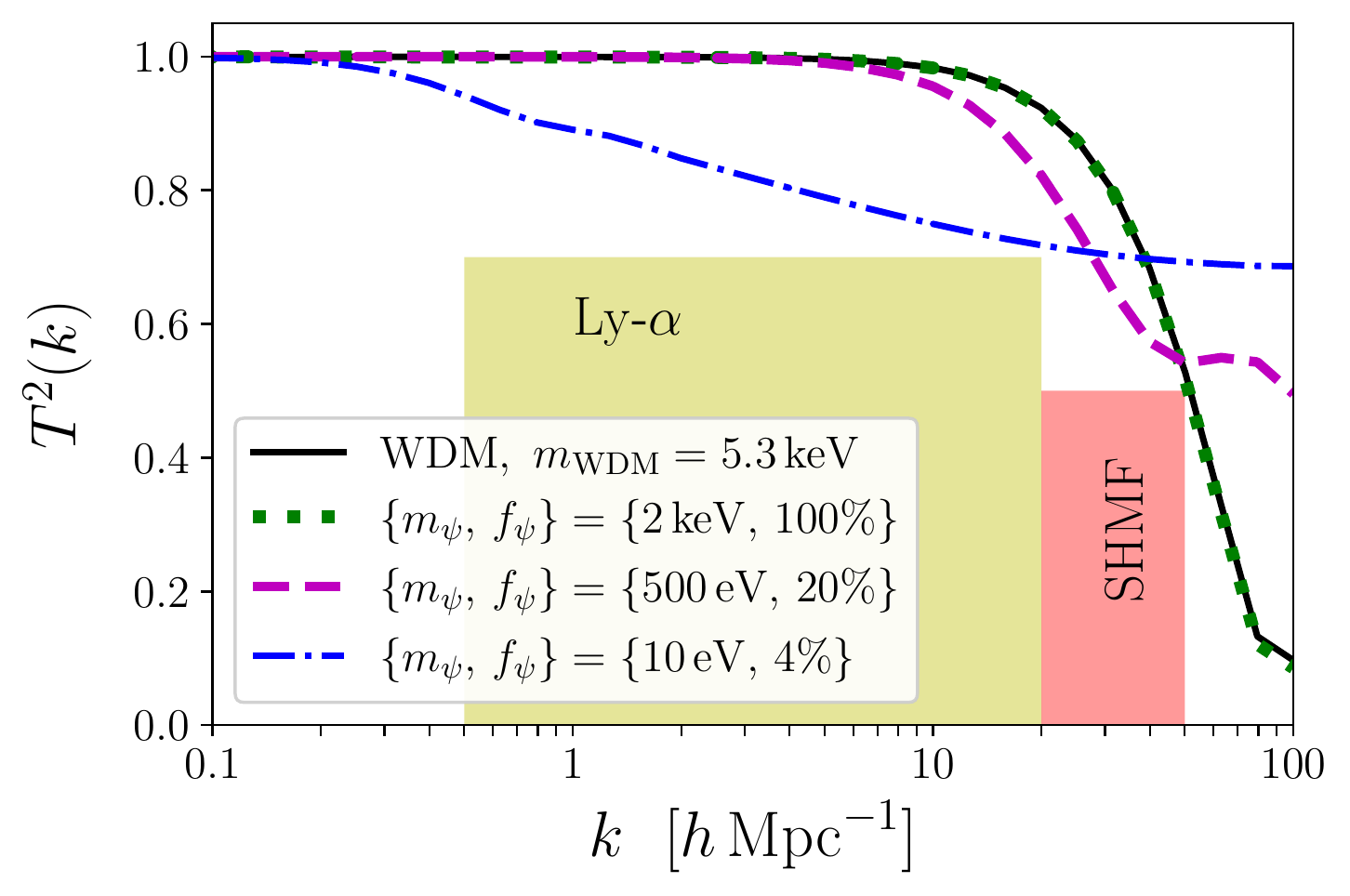}
\caption{The transfer function $T^2(k)$, normalized as described in the text. Black line: warm DM model with $m_{\rm WDM} = 5.3 \kev$. Green dotted line: degenerate fermions with $\{ m_{\psi}, f_{\psi} \} = \{2 \kev, 1 \}$. Magenta dashed line:  degenerate fermions with $\{ m_{\psi}, f_{\psi} \} = \{500 \ev, 20\% \}$, Blue dot-dashed line: degenerate fermions with $\{ m_{\psi}, f_{\psi} \} = \{ 10\ev, 4\% \}$. Transfer functions that pass through the yellow (red) shaded regions are in violation of Ly-$\alpha$ observations (subhalo counts).}
\label{fig:pk}
\end{center}
\end{figure}

We set bounds based on two separate criteria. The Lyman-$\alpha$ forest is sensitive to wavenumbers from $0.5<k{\rm Mpc}/h<20$, and we estimate the corresponding constraint to be $T^2(k  < 20h/{\rm Mpc})\geq 0.7$. This is chosen by looking at the $T(k)$ of WDM scenarios studied in~\cite{Murgia:2017lwo} that pass the Lyman-$\alpha$ constraint. This is supported by the fact that the deviation of the 1D power spectrum in our model mainly comes from the highest $k$-modes $\approx 20h/$Mpc using the data in~\cite{Murgia:2017lwo}. The SHMF is informed by the fact that the smallest satellite galaxies have $k\approx 50h/$Mpc. We estimate the bound from the SHMF to be $T^2(k < 50h/{\rm Mpc})\geq 0.5$. This is chosen based on the power spectrum of the WDM model that passes the bound obtained in~\cite{DES:2020fxi} from DES~\cite{DES:2018gui} and Pan-STARRS1~\cite{chambers2019panstarrs1} data. We represent these constraints as shaded regions in Fig.~\ref{fig:pk}. In Fig.~\ref{fig:pk}, we also show a few $T^2(k)$ examples with different $\{ m_{\psi}, f_{\psi} \}$ that pass these constraints, and we also compare the results to a scenario that only contains WDM and has $m_{\rm WDM}=5.3$~keV, corresponding to the WDM constraint obtained in~\cite{Irsic:2017ixq}. (Different values of $\Lambda$CDM parameters lead to an $\sim \cO(10\%)$ different value for this bound \cite{DES:2020fxi}.) The smallest wavenumber where the suppression is significant depends on the transition redshift $z_t$.

In \Fig{exclusion} we show the resulting exclusion region due to Lyman-$\alpha$ and SHMF constraints in cyan. For $f_\psi=1$, we are able to constrain fermion masses up to $m_\psi=2$ keV due solely to the impacts of degeneracy pressure on structure formation in the early Universe, particularly on the modes of size $k=50h$/Mpc. 
Our bounds asymptote to approximately $f_\psi \lesssim 3\%$ for $m_\psi \lesssim 1$ eV. We are able to constrain down to fractions as low as $3\%$ despite having constraints only at the level of $T^2(k  < 20h/{\rm Mpc})\geq 0.7$ (see, e.g. the blue dot-dashed in Fig.~\ref{fig:pk}), as these degenerate fermions add non-negligibly to the total energy density of the Universe during the time of cosmic structure formation, and change the rate at which density perturbations grow~\cite{Lesgourgues:2006nd}. 

We emphasize that all of our results have been derived in the zero-$T_\psi$ limit, and are only {\it strengthened} if the dark sector has a nonzero temperature, which could be understood intuitively as follows:
a degenerate population of fermions, which occupy the lowest-available states, contains fewer relativistic particles before and during the structure formation era compared to a fully thermalized WDM scenario. Therefore fermionic WDM produces larger values of $\DNeff$ and modifies the matter power spectrum more than degenerate populations of fermions. Allowing lower temperature weakens the bounds from both $\DNeff$ and structure formation on $m_\psi$ until it saturates our constraint in \Fig{exclusion}. We discuss these behaviors in more detail in Appendix~\ref{ap:distributions}.

Our results thus give the minimal mass of fermionic particles allowed with a given flavor number, requiring only their co-moving number density to be conserved since the beginning of BBN 
to the present time. When $f_{\psi} = 1$, our estimate of the Lyman-$\alpha$+SHMF bound shows that fermionic DM that are stable and freeze out before BBN should be heavier than $\approx2 \kev (N_f=1)$, irrespective of their thermal history or phase-space distribution.

~\\
{\bf Local implications.} Besides changing DM structure larger than $\mathcal{O}(20)$~kpc, degeneracy pressure from $\psi$ particles can also modify the core of dwarf galaxies significantly if $f_\psi\approx 1$~\cite{2013MNRAS.429L..89A,Randall:2016bqw}. 
Updated constraints from a survey of the density profiles of MW dwarf satellites give the constraint $m_\psi \geq 130 \ev$ \cite{Alvey:2020xsk}.
This bound is nevertheless weaker than the Lyman-$\alpha$ and SHMF constraints we have obtained by considering the matter power spectrum at smaller wavenumbers.
A related application of this analysis is as an explanation of the core density profiles of dwarf galaxies~\cite{Destri:2012yn, Domcke:2014kla, Alexander:2016glq}. For this purpose, light fermionic DM should be in the range $70\leq m_\psi/\!\ev\leq 400$~\cite{Randall:2016bqw}, which however has been excluded by our constraints.
The $\{ m_{\psi}, f_{\psi} \}$ bound may be relaxed by increasing the number of flavors $N_f$ \cite{Davoudiasl:2020uig}: as shown in Eq.~(\ref{eq:zt}), there is a parameter degeneracy $m_{\psi}\propto N_{f}^{-1/4}$ between $m_\psi$ and $N_f$ when fixing $z_t$ and $f_\psi$ to determine cosmological observables.
However, as derived in~\cite{Randall:2016bqw}, the core radius in the degenerate fermionic DM scenario also scales as $m_\psi N_f^{1/4}$. 
Thus, relaxing our bounds by increasing the number of flavors also reduces the degeneracy pressure in dwarf galaxies. We conclude that repulsion from fermion degeneracy as an explanation of the core-cusp problem is incompatible with matter power spectrum measurements, if the model satisfies our initial assumptions.

Similarly, the Fermi velocity of degenerate fermions will modify features of the $\psi$ population in the overall MW halo, which has an overdensity of DM particles of size $\delta_{\rm MW} = \frac{0.3\gev\!/\!\cm^3}{\Omega_{\rm DM} \rho_c} \simeq 2\tenx5$. Assuming a local $\psi$ density of $\rho_\psi=f_{\psi} \Omega_{\rm DM} \rho_c$, the nonrelativistic Fermi velocity $v_F(z=0) \simeq (m_\psi/\ev)^{-4/3}f_\psi^{1/3}  \times 198\km\!/\!\s$ will exceed the MW escape speed, $v_{\rm esc,\,MW}\simeq  540 \km\!/\!\s$ if $\psi$ is sufficiently light. If so, these particles would not be gravitationally bound to the MW's halo and the actual local energy density $\rho_\psi^{\rm local}$ would not exceed $\rho_\psi$. The situation corresponds to the region on the left of the magenta dashed line in~\Fig{exclusion}. In the region to the right of the magenta dashed line (with larger $m_\psi$), the Pauli degeneracy of $\psi$ will not entirely prevent $\psi$ particles from accumulating in the MW. 
Thus, to the right of the dashed magenta line, we anticipate that there is a local overdensity of $\rho_\psi^{\rm local} > \rho_\psi$, though not necessarily as large as $\delta_{\rm MW}$.

At the dotted magenta line, the Fermi velocity $v_F^{\rm local}=v_{\rm esc,\,MW}$ for $\rho_{\psi}^{\rm local}=0.3 f_\psi \gev\!/\!\cm^3$; thus to the right of this line one can obtain a local $\psi$ overdensity consistent with the local MW DM overdensity $\delta_{\rm MW}$ while satisfying $v_{F}<v_{\rm esc,\,MW}$, which suggests the $\psi$ particles have roughly the same velocity distribution as the CDM. In between the dashed and dotted lines, we can examine two cases to understand the $\psi$ phase-space distribution. In the first case, we require the Fermi velocity to be below $v_{\rm esc,\,MW}$, implying that the local $\psi$ overdensity must be smaller than $\delta_{\rm MW}$. Therefore, the $\psi$ will have a different overdensity distribution from the local CDM distribution, and the velocity distribution may or may not be skewed relative to the virial distribution of the local CDM. 
On the other hand, we note that it is in fact allowed to have a small portion of $\psi$ particles with velocity above $v_{\rm esc,\,MW}$ in the MW~\cite{Kurinsky:2020dpb}. Thus we can have $\rho_{\psi}^{\rm local}  \simeq 0.3 \, f_\psi \gev/\cm^3$ in between the magenta lines, but the $\psi$ will have a velocity distribution with $v_F > v_{\rm esc,\,MW}$. 
The higher velocity due to the degeneracy pressure would make detecting such light DM particles easier than if they had the virial velocity distribution \cite{Kurinsky:2020dpb}, potentially opening up new possibilities for detector materials \cite{Coskuner:2021qxo, Blanco:2021hlm}. In addition, due to the high velocity near $v_{\rm esc,MW}$, the phase space density of particles could be suppressed in the process of MW formation, which can potentially be revealed by $N$-body simulations.
\\
\\
{\bf Models for $\psi$ production.}
We discuss two possible realizations of populating $\psi$ in a degenerate state in the early Universe. The first possibility could be that:
\begin{itemize}
\item $\psi$ particles are generated from the decay of a coherently oscillating scalar field. The scalar field could be the inflaton that generates a fermionic preheating~\cite{Greene:1998nh, Chen:2015dka},  or another scalar field that later decays~\cite{Bjaelde:2010vt}. 
\end{itemize}
In the case of the fermionic preheating through the CP-even inflaton field $\phi$, one assumes a potential $V(\phi) = \frac{1}{2}m^2_\phi\phi^2$ and a Yukawa coupling $y\phi\bar{\psi}\psi$. Since we are considering fermionic particles with $m_\psi \leq \kev \ll m_\phi$, we can approximate its parametric resonance production using the estimates for a massless fermion~\cite{GarciaBellido:2000dc}. Parametric resonance production of $\psi$ will lead to a nearly degenerate Fermi spectrum with momenta stochastically filling a sphere of radius $p_F\sim c^{1/4} m_\phi$ where $c = \frac{y^2 \phi^2_0}{m^2_\phi}$ with $\phi_0$ the inital displacement of $\phi$. For $c \sim 1$, an average fraction $\eta \approx 0.4$ of the states below $p_F$ could be filled up~\cite{Greene:1998nh}. Fermions can also have a parametric resonance production by coupling to an oscillating CP-odd scalar field $\phi_A$. Ref.~\cite{Adshead:2015kza, Adshead:2018oaa} consider such a production of fermion $\psi$ that has a derivative coupling to the axion field. When $m_\psi \sim m_{\phi_A}$, the average  occupation probability could reach $\eta\approx 0.5$~\cite{Adshead:2015kza}. 
Since our bound depends on the combination of $\eta\, m_\psi^4$ for a fixed $\rho_\psi$, as shown in Appendix~\ref{ap:distributions}, the bound on $m_{\psi}$ therefore gets $\eta^{-1/4}\approx 1.2$ times stronger in these scenarios.
\\

Another possibility for obtaining a population of degenerate $\psi$ particles is
\begin{itemize}
\item $\psi$ particles form composite states at early times. 
\end{itemize}
Suppose that the $\psi$ particle is coupled to a scalar $\phi$ with a time-varying mass, $m_\phi$, which mediates an attractive Yukawa force. When the $\psi$ particles are at high density, and their spacing falls below the inverse of $m_\phi$, the Yukawa force can create bound states of the fermions. If the bound states are composed of even numbers of $\psi$ particles, these composite states can then achieve high densities without experiencing degeneracy pressure. The redshift dependence of the energy density of this $\psi\psi$ condensate will depend on the shape of the $\phi$ potential, but can scale like nonrelativistic matter at early times. If $m_\phi$ increases at late times, the Yukawa force can be weakened, and the bound states decay to a high density of individual $\psi$ particles. Models with more complicated dark sectors giving rise to effective phonon-like forces \cite{Berezhiani:2015bqa}, or a model with a nonabelian gauge group \cite{Alexander:2018fjp, Alexander:2020wpm}, can also create bound states in the early Universe. Before BBN starts, we expect these bound states to decay to a high density of individual $\psi$ particles if a mechanism leading to massive gauge bosons becomes effective.
\\ 
\\ 
{\bf Conclusions.} Fermions cannot reach arbitrarily high density without obtaining significant kinetic energy. In this paper, we explored the cosmological implications of degeneracy-induced Fermi momentum and derived qualitatively new bounds on the fermionic DM at zero temperature. The Fermi momentum can cause the dark matter to behave as extra radiation density in the early Universe, thereby contributing to $\DNeff$ at BBN and CMB. It can also prevent the dark matter from aiding in the growth of structure until too low of a redshift, resulting in a suppression of the matter power spectrum.

The dark sector may be richer than just a single particle species. If this is the case, then multiple species contribute to the measured value of $\Omega_{\rm DM}$. Parameterizing the contribution of a particle $\psi$ as the fraction $f_\psi$, we can establish bounds throughout the $m_\psi-f_\psi$ parameter space. Our considerations of the matter power spectrum lead to constraints that can be extended down to $f_\psi$ as small as 3\% for masses $m_\psi \lesssim 1\ev$, and contributions to $\DNeff$ allow constraints on values of $f_\psi$ as small as $2\tenx{-5}$ for $m_\psi \lesssim 0.1\,$meV. For $f_\psi = 1$ we improve existing bounds on $m_\psi$ by an order of magnitude, to $m_\psi=2$ keV. Moreover, we have shown that the local phase space density of DM particles can differ from the MW's virial distribution, and may be suppressed for $m_\psi \lesssim 10 \ev f_\psi^{1/4}$. Near the boundary of this region, it is possible that these particles have an interesting, high-velocity distribution that may be probed in upcoming experiments.

\acknowledgments{
Fermilab  is  operated  by  Fermi  Research  Alliance, LLC under contract number DE-AC02-07CH11359 with the United States Department of Energy. The work of NC at the University of Chicago is supported by the U.S. Department of Energy grant DE-SC0013642. The work of YT is supported by the NSF grant PHY-2014165.
}

\bibliography{fermiversebib}

\appendix

\section{Degenerate Fermion Thermodynamics}
\label{ap:thermo}

\begin{figure}[t]
\begin{center}
\includegraphics[width=0.496\textwidth]{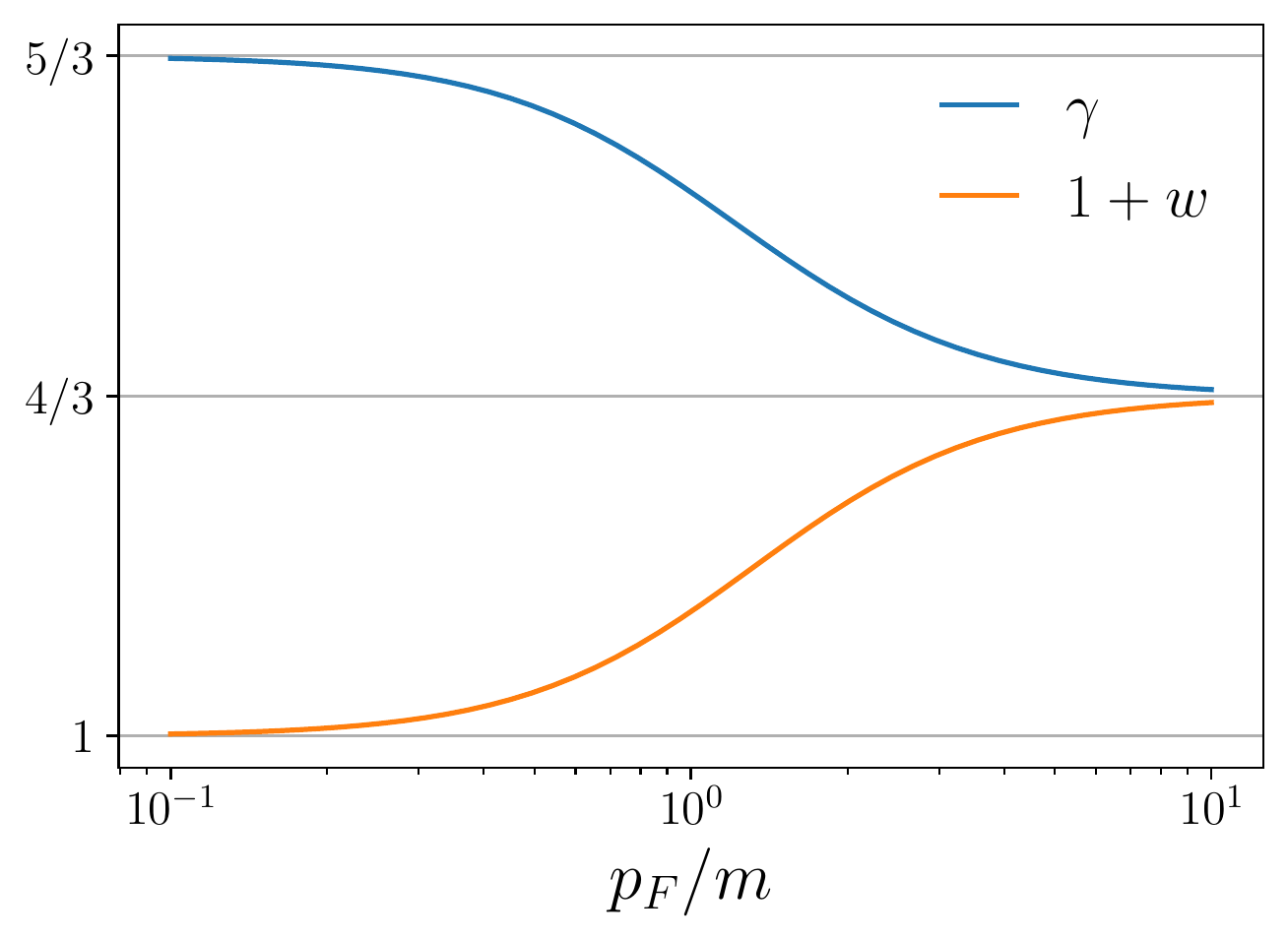}
\caption{The adiabatic index $\gamma$ and the parameter that controls the redshift behavior, $1+w$, as a function of $x=p_F/m$.}
\label{pressure}
\end{center}
\end{figure}
For any instantaneous time with a given $p_F$, using Eqs.~1.7, 1.20, 1.22 from Ref. \cite{Jaffe-notes}, we have
\alg{ \label{defs}
n_\psi = \frac{g_\psi\, p_F^3}{6\pi^2},& \qquad \rho = g_\psi \int_0^{p_F} \frac{d^3 p \, E}{(2\pi)^3},
\\ P = \frac{\rho' n_\psi- \rho n_\psi'}{n_\psi'},& \qquad \gamma =  \frac{n_{\psi}^2}{n_\psi'^2} \frac{\rho'' n_\psi' - \rho' n_\psi''}{n_\psi \rho' - n_\psi' \rho}
}
where the $n-p_F$ relation can be taken to define $p_F$, $E = \sqrt{m^2+p^2}$ is the energy of the particle, and $'$ denotes $d/dp_F$.
As we are using the {\it entire} energy $E=\sqrt{m^2+p^2}$ to calculate $\rho$, the equation of state is simply $w=P/\rho$.

We have analytic results for $\rho$, $P$, and $\gamma$:
\alg{
\label{eq:degexp}
\frac{\rho}{m^4} &= \frac{g_\psi}{8\pi^2} \bL (2x^3+x)\sqrt{1+x^2} - {\rm asinh}(x) \bR,
\\ \frac{P}{m^4} &= \frac{g_\psi}{24\pi^2} \bL (2x^3-3x)\sqrt{1+x^2} + 3\,{\rm asinh}(x) \bR,
\\ \gamma &= \frac{8x^5/3}{2x^5-x^3-3x+3\sqrt{1+x^2}\,{\rm asinh}(x)},
\\ \frac{P}\rho &= \frac{8x^3\sqrt{1+x^2}/3}{(2x^3+x)\sqrt{1+x^2}-{\rm asinh}(x)} -1,
}
where $x=p_F/m$ and ${\rm asinh}(x) = \ln(x+\sqrt{1+x^2})$.
These expressions are plotted in \Fig{pressure}. The adiabatic index asymptotes to 5/3 when $x \to 0$, but $w$ asymptotes to 0. At high $p_F$, both $\gamma$ and $1+w$ tend to 4/3. In the expanding Universe, the quantities above would depend on the instantanuous redshift. Since the second Friedman equation is written $d\rho/d a= -3H(\rho + P)$, $\rho$ scales as $\rho \propto a^{-3(1+w)}$ for slowly varying $w$. Therefore, we see that the $\psi$ fluid redshifts like radiation for $p_F \gtrsim m_\psi$ and redshifts like cold matter for $p_F \lesssim m_\psi$.

\section{Comparison with Other Distributions}
\label{ap:distributions}
Fermionic DM particles that are relativistic before thermal decoupling follow a momentum distribution \begin{equation}
\label{eq:distribution_0}
    f(q) = \frac{\eta}{1 + e^{(q-\mu)/T_\psi}}\,.
\end{equation}
Here $q$ is the co-moving momentum, $T_\psi$ is the dark sector temperature at $z=0$. $\eta$ is a factor that depends on the thermal history of the dark sector such as the occupation probability of the fermion energy states. In literature, the  chemical potential is usually set to $\mu = 0$~\cite{Colombi:1995ze, Baur:2015jsy} as in the discussion of warm DM models~\cite{Baur:2015jsy}.

The physical momentum of DM is given by: $p = q/a$. The number density, energy density and pressure for each flavor are given by:
\begin{eqnarray}
\label{eq:distribution}
a^3 n_\psi(a) &=& \frac{ g_\psi\eta}{(2\pi)^3}\int d^3 q f(q) \\
a^3\rho(a) &=& \frac{g_\psi \eta}{(2\pi)^3}\int d^3 q f(q) \sqrt{m_{\psi}^2 + (\frac{q}{a})^2} \notag\\ 
&=&\frac{m^4_{\psi} g_\psi \eta}{(2\pi)^3}\int d^3 q f(q m_\psi) \sqrt{1 + (\frac{q}{a})^2} \notag
\end{eqnarray}
\begin{eqnarray}
a^{3}P(a) &=& \frac{g_\psi \eta}{(2\pi)^3}\int d^3 q f(q) \big(\frac{q}{a}\big)^2 \frac{1}{3\sqrt{m_{\psi}^2 + (\frac{q}{a})^2}} \notag\\ 
&=& \frac{m^4_{\psi} g_\psi \eta}{(2\pi)^3}\int d^3 q f(q m_\psi) \big(\frac{q}{a}\big)^2 \frac{1}{3\sqrt{1 + (\frac{q}{a})^2}}. \notag
\end{eqnarray}
Then we have $\rho_\psi = N_f \rho(a=1)$. For $\mu = 0$, DM models with the same $(\rho_\psi,\frac{m_\psi}{T_\psi})$ have identical $\rho(a)$ and $P(a)$, which lead to the same contribution to $\DNeff$ and the same structure formation process. Fixing $\frac{m_\psi}{T_\psi}$ makes $\rho_\psi\propto \eta N_f m_{\psi}^4$ and leads to a degeneracy $m_{\psi}\propto N_f^{-1/4}$ in the power spectrum constraint.

If $\mu\neq 0$, $(\rho_\psi,\frac{m_\psi}{T_\psi})$ no longer specify $\rho(a)$ and $P(a)$. Since the $\DNeff$ and the matter power spectrum are mainly sensitive to  $a_t=(1+z_t)^{-1}$ for particles transiting from relativistic radiation to non-relativistic matter, models with the same $a_t$ and  $\rho_\psi$ produce similar $\DNeff$ and matter power spectra. From numerically solving the integrals in Eq.~(\ref{eq:distribution}), one can show that in order to keep a similar $\DNeff$ and structure formation bound by fixing $(a_t,\rho_\psi)$, lowering $T_\psi$ would require an increase in $\mu/T_\psi$ and a decrease in $m_{\psi}$. This means when lowering the temperature of fermionic DM, the bounds become weaker and asymptote to the zero temperature bounds we derived. In the extreme case with $T_\psi \rightarrow 0$, we thus expect the bounds give the minimal mass allowed for fermionic dark matter, consistent with the intuition discussed in the main text.

For degenerate fermionic state in the early Universe, the momentum-space distribution in the co-moving frame can be written as:
\begin{equation}
\label{eq:fdeg}
    f(q) = \theta(q_F - q)
\end{equation}
which could be realized by choosing $\eta = 1, \mu = q_F$ and $T_\psi \ll q_F$ in \Eq{eq:distribution_0}.
For the following discussions, we keep the dependence on $\eta$ explicitly. The number density at $z=0$ is then calculated as:
$n_\psi(z=0) = \eta\frac{g_\psi q^3_F}{6\pi^2}$ where we identify $q_F = p_F(z=0)$. The time-dependent $P(a)$ and $\rho(a)$ reduces to that in \Eq{eq:degexp} by identifying $p_F = q_F /a$, and thus degenerate fermion models with the same $(\rho_\psi, \frac{q_F}{m})$
would have identical $\rho(a)$ and $P(a)$, predicting the same $\DNeff$ and large scale structure. Fixing $\frac{q_F}{m}$, it follows that $\rho_\psi \propto \eta N_f m^4_\psi$, and hence the bounds on $m_\psi$ also scale as $N^{-1/4}_f$.
\end{document}